# Characterization of Alkali Metal Dispensers and Non-Evaporable Getter Pumps in Ultra-High Vacuum Systems for Cold Atomic Sensors

David R. Scherer, David B. Fenner, and Joel M. Hensley

Physical Sciences Inc., 20 New England Business Center, Andover, MA 01810

A glass ultrahigh vacuum chamber with rubidium alkali metal dispensers and non-evaporable getter pumps has been developed and used to create a cold atomic sample in a chamber that operates with only passive vacuum pumps. The ion-mass spectrum of evaporated gases from the alkali metal dispenser has been recorded as a function of dispenser current. The efficacy of the non-evaporable getter pumps in promoting and maintaining vacuum has been characterized by observation of the Rb vapor optical absorption on the $D_2$ transition at 780 nm and vacuum chamber pressure rate of rise tests. We have demonstrated a sample of laser-cooled Rb atoms in this chamber when isolated and operating without active vacuum pumps.

## I. INTRODUCTION

Compact, transportable ultrahigh vacuum (UHV) chambers that operate with only passive vacuum pumps are required for the development of fieldable, miniature precision sensors based on a sample of laser-cooled alkali atoms. The preparation of an ensemble of laser-cooled atoms in such a UHV chamber forms the starting point for cold atomic sensors[1] such as atomic clocks, atomic magnetometers, and inertial sensors that rely on atom interferometry. Compact, transportable ultra-high vacuum suitcases with non-evaporable getter (NEG) pumps have been developed for the transportation of hazardous or sensitive materials[2] and compact vacuum chambers that use a NEG pump in conjunction with a turbo pump[3] or ion pump[4] have been reported. However, there has been, to our knowledge, no demonstration of laser cooling in a compact, glass UHV chamber that operates with only NEG pumps intended for transportable cold atomic sensors.

The preparation of a cold atomic sample relies on laser cooling of an atomic vapor. This process is performed in a UHV ($P < 1 \cdot 10^{-9}$ Torr) chamber with a controllable source of alkali metal atoms (e.g. Li, Na, K, Rb, Cs) such as an alkali metal dispenser,[5,6,7] a cold finger,[8] or an atomic beam source and Zeeman slower.[9] An alkali atom density on the order of $n \sim 10^9$ cm$^{-3}$ is produced to ensure that a sufficient number of atoms can be laser cooled in a magneto-optical trap (MOT) while maintaining the chamber background (non-alkali) pressure at less than $10^{-9}$ Torr to minimize collisions with background gases. The main chamber body is made of glass to allow for the interrogation of alkali atoms in the body of the chamber with incident laser light. For atomic state preparation on magnetically sensitive levels or for operation as an atomic magnetometer, the chamber and vacuum pumps should be made of non-magnetic materials to minimize Zeeman energy shifts, stray magnetic fields, and eddy currents.

To reduce the size of vacuum chambers for cold atomic systems to a volume on the order of ~ 1 cm$^3$, the large active vacuum pumps common in research and industrial settings must be replaced with compact, passive pumps. We have developed a prototype version of such a system, characterized its vacuum properties and demonstrated laser cooling of Rb atoms in this isolated chamber without active vacuum pumps.

The vacuum system consists of a 500 cm$^3$ volume glass chamber with side arms that contain Rb alkali metal dispensers (AMDs) and NEGs. Our Rb source is a commercially available alkali-metal dispenser (model Rb/NF3.4/12, SAES Getters). The AMD is a mixture of an alkali-metal chromate (RbCrO$_4$) with a reducing agent. The reducing agent (SAES St101) composition is 84% Zr and 16% Al, resulting in a getter material that is able to irreversibly sorb chemically active gases created during the reduction reaction, preventing them from contaminating the chamber while releasing Rb vapor into the chamber. Thus, this Rb dispenser is able to produce a controllable source of Rb vapor by evaporation during ohmic heating. Heater operation at a current in the range of 3 to 7 A heats the Rb dispenser to a temperature range of ~400-800 °C.[10]

The NEG pumps, developed commercially by SAES Getters (model St172), are a blend of zirconium powder and SAES St707 compound, which is an alloy of Zr, V, and Fe.[11,12] These materials are mixed together and sintered at high temperatures to produce a material with high porosity and large effective surface area. In order to remove the thin oxidation layer present after manufacturing, the initial activation, or degassing, procedure requires operation at 900°C for 10 minutes under a vacuum of 1 mTorr or less. After activation, the NEG works by sorbing gases such as CO, $CO_2$, $O_2$, and $N_2$ and forming stable chemical compounds such as zirconium carbide, zirconium oxide, or zirconium nitride.[13] These compounds do not decompose at temperatures below 1000°C, making their sorption irreversible. $H_2$ can diffuse into the bulk NEG material, resulting in reversible sorption. Water vapor is sorbed via dissociation into $O_2$ and $H_2$ on the NEG surface, and hydrocarbons such as $CH_4$ are sorbed by thermal activation only above 300°C. Because the NEG does not pump Rb, we treat the pumping capacity of the NEG as independent of the Rb vapor density. Similarly, the production of Rb vapor density from the AMD is not affected by the presence of the NEG.

After activation, the gettering rate, or pumping rate, of the NEG depends on the gas species being pumped as well as the quantity sorbed. For large quantities of sorbed gases, the pumping rate for gases such as $H_2$ and CO by the NEG pump decreases as the NEG surface saturates. Our glass chamber, which has a volume of 500 $cm^3$, operates at a pressure below $10^{-9}$ Torr, where the pumping rate of the NEG pump is at its maximum. The manufacturer-specified pumping rate for $H_2$ is ~2 L/s and for CO is ~0.2 L/s at room temperature, and both of these pumping rates increase by approximately a factor of two for NEG operation at an elevated temperature of 200° C due to increased diffusion into the bulk NEG material. Although the NEG does not pump the inert gas He, He permeation into the chamber can be mitigated by using a transparent chamber material with low He permeability, such as the glass ceramic Zerodur.

## II. EXPERIMENTAL APPARATUS AND PREPARATION

A schematic of our vacuum chamber is shown in Figure 1(a). The glass chamber is a Pyrex glass cell with three side arms. One side arm contains the Rb AMD and two side arms contain one NEG pump each. The side arms are oriented to avoid line-of-sight contamination of the NEGs by evaporated gases from the AMDs. This glass chamber is connected, via a UHV gate valve, to a larger vacuum chamber for initial pumping and characterization. The square portion of the glass cell, used for optical access, is 10.5 cm x 3.8 cm x 3.8 cm, the entire glass chamber has a volume of ~500 $cm^3$ and an interior surface area of ~600 $cm^2$. Both the AMDs and the NEGs are connected to stainless steel electrical leads accessible outside the cell via a tungsten pin-press spot-welded to nickel leads in the interior of the chamber. The glass chamber is connected to a glass-to-metal seal and a 2.75" Conflat flange. A gate valve (MDC GV-1500-M-P) with a metal bonnet seal is used in between the glass chamber and the main body of the vacuum system, which serves to isolate the glass chamber from the active pump and simulate the effects of permanently removing the glass chamber from the rest of the vacuum system.

The gate valve is connected to a six-way cross. The arms of the cross contain a venting valve for dry $N_2$ purging, a residual gas analyzer (ExTorr XT100 RGA), a quartz viewport, an ion gauge (MKS 909AR-21) and a turbo pump (Pfeiffer Balzers 56 L/s pump) backed by a rotary roughing pump.

With the gate valve closed and the glass chamber unattached, the vacuum system was baked at 180° C for 40 hours to ensure cleanliness of the stainless steel components. After completion of the initial bakeout, the glass chamber was attached and the gate valve was opened. A second bakeout at 200° C for 40 hours was performed to ensure the cleanliness of the glass chamber. At the completion of the second bakeout, activation, or degassing, of the Rb dispensers and NEG pumps was performed. The NEG pumps were activated first in order to fully desorb the surface oxide layer that exists upon shipment and allow the NEGs to pump the evaporated gases that occur during the Rubidium dispenser activation. Both NEGs were activated at the same time to avoid cross-contamination. To activate the NEGs, a current of 3.3 A was operated through both NEGs simultaneously for 10 minutes. During activation, the pressure in the chamber rose to a maximum of $10^{-5}$ Torr after ~1 minute and then dropped to $9 \cdot 10^{-7}$ Torr after 10 minutes. Activation of the Rb dispensers was performed by increasing the current slowly up to the activation current of 4.7 A over a total of 27 minutes. Operation of the Rb AMD at a current above ~4 A causes a small amount of alkali metal adsorption on the interior of the glass sidearm immediately next to the AMD. We have not observed Rb adsorption elsewhere on the interior of the glass chamber after months of operation.

## III. RUBIDIUM DISPENSER AND VACUUM CHARACTERIZATION

In order to characterize the non-Rb (contaminant) gases that evaporate from the AMD as a result of its elevated temperature, the Rb dispenser was operated at an elevated current, the chamber was allowed to stabilize for ~30 minutes, and an RGA ion-mass spectrum was recorded. As Rb is highly reactive, most of the Rb atoms that evaporate from the dispenser are expected to adsorb to the Pyrex chamber walls with high probability (sticking coefficient ~1) rather than arrive at the RGA, and Rb atoms were not observed on the RGA mass spectrum.

A plot of the RGA mass spectrum from mass 1 amu to 110 amu is shown in Figure 2 for increasing values of Rb dispenser current. The largest peak in the RGA spectrum is at mass 28 ($N_2$, $C_2H_4$, and CO). No doubly-ionized species were observed in the RGA spectrum. For most masses, the partial pressure increases as the Rb dispenser current increases from zero to 4.75 A. However, this is not the case for mass 18 ($H_2O$), mass 32 ($O_2$), or mass 44 ($CO_2$). In the inset in Figure 2, the pressure increase for $CO_2$ (mass 44) is shown. The partial pressure of $CO_2$ reaches a maximum for a dispenser current of 4 A, saturates, and then decreases at higher dispenser currents. A similar local maximum in partial pressure was observed for $H_2O$ and $O_2$. Although contaminant (non-alkali) gas evaporation data for SAES Rb alkali-metal dispensers is not available, K dispensers from the same manufacturer exhibit a similar saturation and decrease in $CO_2$ partial pressure for increased dispenser current.[14] Subsequent to the Rb AMD characterization, the RGA was removed from the chamber and the chamber pressure decreased from $1.9 \bullet 10^{-9}$ Torr to its base pressure of $P = 2.3 \bullet 10^{-10}$ Torr as recorded on an ion gauge. The steady-state base pressure recorded on the ion gauge does not change significantly if the gate valve is open or closed.

After operation and shut-off of the Rb dispenser, the Rb vapor density measured by optical absorption decays quickly, with a 1/*e* lifetime of 5 seconds, independent of the initial Rb density and dispenser operation time. This serves as an indication of the efficacy of the Pyrex chamber walls in acting as a pump for Rb, which is not pumped by the NEG.

With the Rb dispenser off, the efficacy of the NEG pumps in maintaining UHV in the absence of an active pumping mechanism can be inferred from a chamber pressure rate of rise test[4]. In this test, the turbo pump and right-angle valve (Figure 1(b)) were simultaneously turned off and the background pressure as recorded on the ion gauge was monitored as a function of time. With the gate valve closed, the background pressure in the region of the six-way cross (which we consider as indicative of the background pressure in the glass chamber) rose from $10^{-9}$ Torr to $10^{-7}$ Torr in ~9 minutes, indicative of the system leak rate and spin-down time of the turbo pump. With the gate valve open, the time required for the pressure in the entire chamber (glass chamber and six-way cross) to rise from $10^{-9}$ Torr to $10^{-7}$ Torr increased by 17% as measured over multiple trials. This slower rate of rise, despite the increase in volume and surface area of the region being evacuated due to the glass cell, demonstrates that the NEGs offer additional pumping capacity in the range of $10^{-9}$ Torr to $10^{-7}$ Torr.

## IV. RUBIDIUM VAPOR OPTICAL CHARACTERIZATION

The Rb atomic density in the chamber was characterized by on-resonance optical absorption measurements using a setup shown in Fig. 1(b). An external-cavity diode laser was scanned in frequency across the $^{85}$Rb and $^{87}$Rb D$_2$ transitions at a central wavelength of $\lambda_0 = 780$ nm. The on-resonance transmission $T_0$ through the Rb vapor in the chamber was used to calculate the total Rb atomic density $n$ based on the on-resonance absorption coefficient $k_0$ using $T_0 = e^{-k_0 l}$. The on-resonance absorption coefficient $k_0$ is related to the number density of atoms $n$ as $k_0 = \frac{2}{\Delta \nu_D} \sqrt{\frac{\ln(2)}{\pi}} \frac{\lambda_0^2}{8\pi} \frac{g_2}{g_1} \frac{n}{\tau}$, where the Doppler linewidth $\Delta \nu_D$ is 500 MHz, the natural lifetime $\tau$ is 26 ns, and the factor $\frac{g_2}{g_1}$ accounts for the statistical weights of the ground and excited states of the D$_2$ transition[15].

The steady-state baseline-normalized optical transmission through the glass chamber (for five passes through the glass cell having a total interaction length of $l = 19.1$ cm) is shown in Figure 3(a) for increasing values of Rb dispenser current. Based on the measured absorption coefficient, the total Rb number density ($^{85}$Rb + $^{87}$Rb) is calculated and shown in Figure 3(b). A steady-state Rb vapor density as high as $7.3 \cdot 10^9$ cm$^{-3}$ was achieved for a dispenser current of 4.5 A in our chamber. For timescales of less than one hour, the Rb vapor density at a given dispenser current does not depend on whether the gate valve is open or closed.

## V. LASER COOLING

We have demonstrated a MOT of $^{87}$Rb atoms in our glass chamber using a fiber-based laser system. The cooling light is generated from a narrow-linewidth telecom seed laser at 1560 nm that is amplified in a 5 W Erbium-doped fiber amplifier (EDFA) and frequency-doubled in a bulk periodically-poled lithium niobate (PPLN) crystal, generating 550 mW at 780 nm. The cooling light is red-detuned 24 MHz from the $^{87}$Rb $5^2S_{1/2}$ |F=2⟩ → $5^2P_{3/2}$|F′=3⟩ transition. The repump light is generated by using an in-line fiber-based phase modulator before the EDFA that generates sidebands on the laser light, one of which is on resonance with the $^{87}$Rb $5^2S_{1/2}$ |F=1⟩ → $5^2P_{3/2}$|F′=2⟩ transition after the frequency doubler. The 780 nm beam is split into three paths and coupled into three independent polarization-maintaining fibers. Fiber beam expanders then expand the beams to a 20 mm diameter, resulting in a retro-reflected, three-beam MOT configuration. Using 80 mW per beam, an axial magnetic field gradient of ~15 G/cm generated from a pair of magnet coils in anti-Helmholtz configuration, and a Rb AMD current of 3.7 A, we trap ~$10^7$ atoms in the MOT based on the steady-state fluorescence as measured on a collection photodiode[5].

The dynamic loading of the MOT is described by the rate equation $\frac{dN}{dt} = R - \Gamma N$, where $N$ is the number of atoms in the MOT, $R$ is the loading rate, and $\Gamma$ is the loss rate[8]. The time-dependent solution to this rate equation is $N(t) = \frac{R}{\Gamma}\left(1 - e^{-\Gamma t}\right)$, whereas the steady-state solution yields $N_{SS} = R/\Gamma$. By closing the right-angle valve in Figure 1(b), we are able to effectively eliminate the backing pump for the turbo pump, allowing for a slow increase in overall background pressure in the chamber and constant Rb vapor density due to the AMD. This resulted in an overall background (non-Rb) pressure rise from $1.9 \cdot 10^{-9}$ Torr to $9.5 \cdot 10^{-8}$ Torr over 20 minutes, as measured on the ion gauge. After this increase in chamber pressure, the number of atoms in the MOT decreased by a factor of 9.3, whereas the loss rate $\Gamma$ increased from 1.1 s$^{-1}$ to 9.4 s$^{-1}$, as measured by fitting the time-dependent solution to the MOT loading rate equation to the MOT loading data. This indicates that the loss rate is dominated by collisions due to background (non-Rb) atoms at this pressure.

After closing the gate valve and isolating the glass chamber from the active vacuum pumps, we are able to load a MOT for a timescale on the order of several hours with the Rb AMDs on, demonstrating that an isolated, pre-evacuated chamber with no active vacuum pumps can be used for the preparation of laser-cooled atomic samples.

# VI. CONCLUSIONS

As researchers continue to reduce the size of sensors and devices based on cold and ultracold atoms, the size of the glass UHV chamber, or physics package, will eventually approach the centimeter scale or less. For the development of a transportable[16] or miniature[17] cold-atom sensor in remote situations outside of the laboratory, the employment of a passive pumping system, such as what we have developed here, will be required. In such a system, the loading time of the MOT itself, rather than a vacuum gauge, may serve as an indication of the chamber pressure.

Our demonstrations show that the Rb dispensers can act as a controllable source of Rb vapor density in the range of $10^8 - 10^{10}$ cm$^{-3}$, which is used to load a sample of cold atoms in the MOT. Our measurements reveal that the NEG pumps provide sufficient independent pumping capacity to maintain UHV pressures for several hours in an isolated chamber of volume ~500 cm$^3$ that has been previously evacuated by a turbo pump, allowing for the demonstration of a MOT in a chamber with no active vacuum pumps. For use in a miniature, transportable, cold-atom sensor in situations outside of the laboratory, the glass chamber could be detached from the main vacuum chamber body by a glass pinch-off procedure subsequent to initial bakeout and activation of the NEGs and AMDs.

## ACKNOWLEDGEMENTS
This work was funded by NAVAIR under contract N68335-10-C-0509. The authors gratefully acknowledge experimental assistance from Robert Krech and Jan Polex.

## VII. REFERENCES

[1] J. Kitching, S. Knappe, and E. A. Donley, IEEE Sens. J. **11**, 1749 (2011).

[2] G. Firpo, and A. Pozzo, Rev. Sci. Instrum. **75**, 4828 (2004).

[3] A. Pozzo, C. Boffito, and F. Mazza, Vacuum **47**, 783 (1996).

[4] C. D. Park, S. M. Chung, and P. Manini, J. Vac. Sci. Technol. A **29**, 011012 (2011).

[5] U. D. Rapol, A. Wasan, and V. Natarajan, Phys. Rev. A **64**, 023402 (2001).

[6] H. J. Lewandowski, D. M. Harber, D. L. Whitaker, and E. A. Cornell, J. Low Temp. Phys. **132**, 309 (2003).

[7] K. L. Moore, T. P. Purdy, K. W. Murch, S. Leslie, S. Gupta, and D. M. Stamper-Kurn, Rev. Sci. Instrum. **76**, 023106 (2005).

[8] C. Monroe, W. Swann, H. Robinson, and C. Wieman, Phys. Rev. Lett. **65**, 1571 (1990).

[9] C. Slowe, L. Vernac, and L. V. Hau, Rev. Sci. Instrum. **76**, 103101 (2005).

[10] M. Succi, R. Canino, and B. Ferrario, Vacuum **35**, 579 (1985).

[11] C. Benvenuti, J. M. Cazeneuve, P. Chiggiato, F. Cicoira, A. E. Santana, V. Johanek, V. Ruzinov, and J. Fraxedas, Vacuum **53**, 219 (1999).

[12] C. Benvenuti, P. Chiggiato, P. C. Pinto, A. E. Santana, T. Hedley, A. Mongelluzzo, V. Ruzinov, and I. Wevers, Vacuum **60**, 57 (2001).

[13] G. L. Saksaganskii, *Getter and Getter-Ion Vacuum Pumps*, (Harwood Academic Publishers, Chur, Switzerland, 1998).

[14] P. della Porta, C. Emili, and S. J. Helier, Proceedings of the Ninth IEEE Conference on Tube Techniques, New York, 1968, pp. 246.

[15] A. C. G. Mitchell and M. W. Zemansky, *Resonance Radiation and Excited Atoms*, (Cambridge University Press, Cambridge, 1934).

[16] D. M. Farkas, K. M. Hudek, E. A. Salim, S. R. Segal, M. B. Squires, and D. Z. Anderson, Appl. Phys. Lett, **96**, 093102 (2010).

[17]V. Shah, R Lutwak, R. Stoner, and M. Mescher, Proceedings of the IEEE International Frequency Control Symposium, Baltimore, 2012.

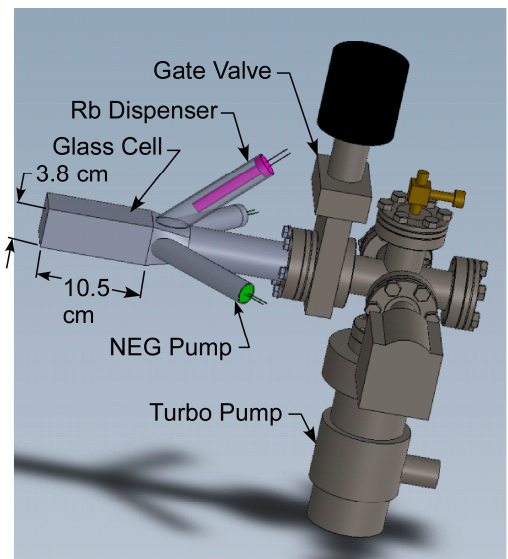

(a)

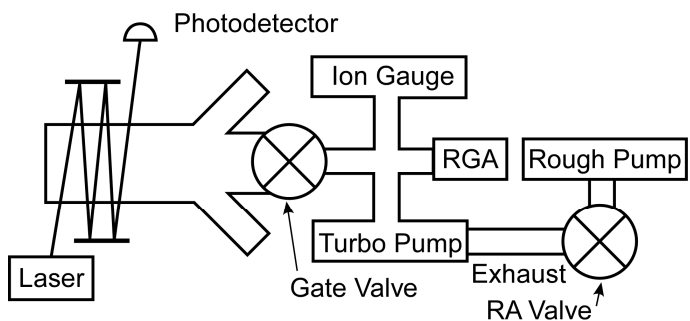

(b)

Figure 1. (a) Schematic of UHV chamber. Glass cell contains Rb dispenser and non-evaporable getter (NEG) pump. Gate valve connects glass cell to main chamber body, which contains a turbo pump and vacuum diagnostic equipment. (b) Schematic of chamber and optical absorption experiment, RGA = residual gas analyzer, RA valve = right-angle valve.

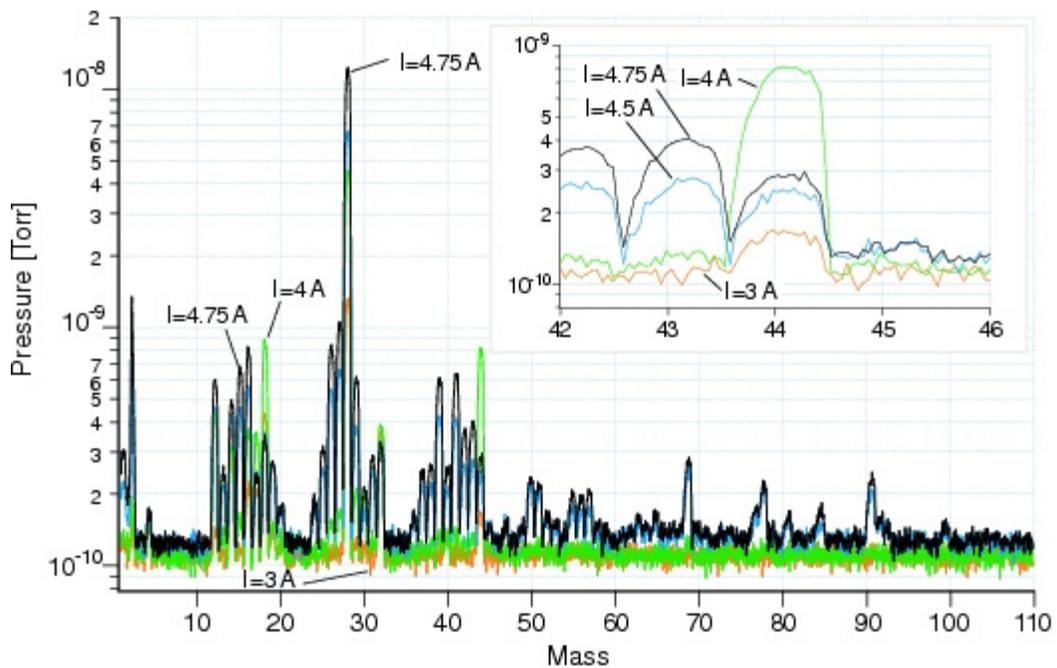

Figure 2. (Color online) RGA mass spectrum up to 110 amu, showing partial pressure (log scale) vs. mass for increasing dispenser currents, as indicated. Inset: RGA mass spectrum from 42 to 46 amu, showing the pressure increase for $CO_2$ (44 amu).

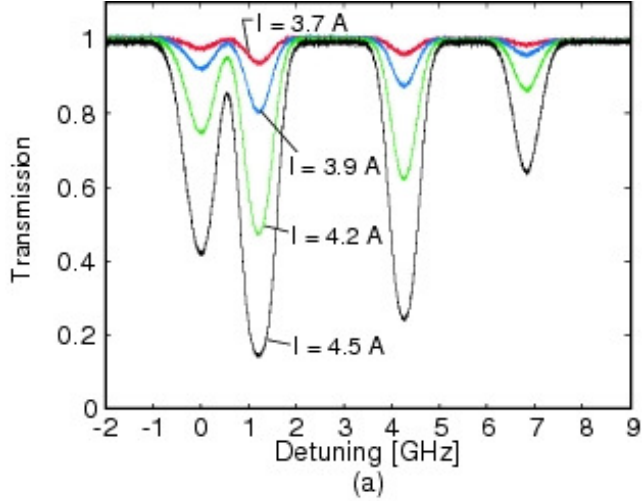

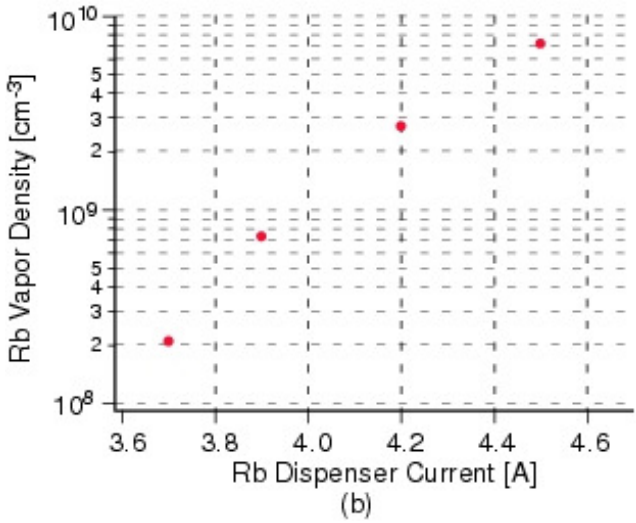

Figure 3. (a) Baseline-normalized optical transmission through glass chamber vs. frequency, shown for increasing values of Rb dispenser current. In order of increasing frequency, the $^{87}$Rb $5^2S_{1/2}$ |F=2> →$5^2P_{3/2}$|F'> transition, $^{85}$Rb $5^2S_{1/2}$ |F=3> →$5^2P_{3/2}$|F'> transition, $^{85}$Rb $5^2S_{1/2}$ |F=2> →$5^2P_{3/2}$|F'> transition, and $^{87}$Rb $5^2S_{1/2}$ |F=1> →$5^2P_{3/2}$|F'> transition are shown, from left to right. Detuning in GHz is shown relative to the $^{87}$Rb |F=2> →|F'> transition. (b) Total measured Rb density (log scale) vs. Rb dispenser current.